# Dirac bands in the topological insulator Bi$_2$Se$_3$ mapped by time-resolved momentum microscopy

Stefano Ponzoni[1,*,§], Felix Paßlack[1, §], Matija Stupar[1], David Maximilian Janas[1], Giovanni Zamborlini[1,*] and Mirko Cinchetti[1].

[1] Department of Physics, TU Dortmund University, 44221 Dortmund, Germany

[§] These authors contributed equally to this work.
*Corresponding authors:
stefano.ponzoni@tu-dortmund.de; giovanni.zamborlini@tu-dortmund.de

## Abstract

We have studied the energy dispersion of the Dirac bands of the topological insulator Bi$_2$Se$_3$ at large parallel momenta using a setup for laser-based time-resolved momentum microscopy with 6 eV probe-photons. Using this setup, we can probe the manifold of unoccupied states up to higher intermediate-state energies in a wide momentum window. We observe a strongly momentum-dependent evolution of the topologically protected Dirac states into a conduction band resonance, highlighting the anisotropy dictated by the surface symmetry. Our results are in remarkable agreement with the theoretical surface spectrum obtained from a GW-corrected tight-binding model, suggesting the validity of the approach in the prediction of the quasiparticle excitation spectrum of large systems with non-trivial topology. After photoexcitation with 0.97 eV photons, assigned to a bulk valence band-conduction band transition, the out-of-equilibrium population of the surface state evolves on a multi-picosecond time scale, in agreement with a simple thermodynamical model with a fixed number of particles, suggesting a significant decoupling between bulk and surface states.

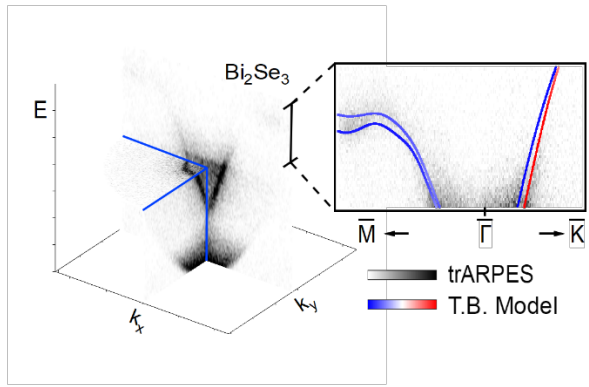

I. INTRODUCTION

At the microscopic level, the functional properties of electronic and opto-electronic devices are determined by the out-of-equilibrium dynamics of their charge-carriers. In semiconductors, for example, generation, recombination, and, in general, manipulation of non-equilibrium electron and hole distributions by external fields enable the control of the macroscopic charge-flow, ultimately leading to the desired functionalities. However, this evolution is not solely determined by the action of the external stimuli but also critically depends on the electronic structure of the material, as well as the elastic and inelastic relaxation pathways available to the excited carriers. Time- and angle- resolved photoemission spectroscopy (trARPES) allows to probe both the out-of-equilibrium electronic structure and the relaxation dynamics of non-equilibrium carrier distributions by measuring the spectral function of a material after the excitation with a femtosecond laser pulse [1,2]. As a result, this technique has been extensively used to investigate the electron dynamics in semiconductors [3–5], the exciton and electron dynamics in organic adsorbates [6,7], the carrier dynamics in 2D materials [8,9] and their unoccupied electronic structure [10], just to name a few.

Recent studies have addressed the unoccupied band structure and the carrier dynamics of topological materials [11]. In particular, tr-ARPES studies on topological insulators (TI) evidenced the possibility to photo-induce strongly out of equilibrium states by acting on the charge balance in their Dirac cone. After the photoexcitation, these two-dimensional Dirac systems show a metallic nature and, at the same time, an extremely long recombination time [12]. The unusual properties of the out of equilibrium carriers in the topologically-protected states, together with their spin-momentum locking [13,14], a hallmark of strong spin-orbit interaction, make this class of materials an ideal candidate for spintronic applications [15], where the functionality of the device is achieved through the control and the manipulation of non-equilibrium spin-distributions together with, or even independently from, the charge carriers' ones. Bismuth Selenide ($Bi_2Se_3$) is an archetypical topological insulator that has been subjected to extensive investigations of its electronic structure [13,14], transport [16–18], and optical properties [19], as well as its non-equilibrium carrier dynamics [11,20–25]. $Bi_2Se_3$ has also been integrated into many experimental spintronic [26,27], optoelectronic [28], and opto-spintronic devices [29]. Despite the impressive extent of experimental findings, the unoccupied states band-structure at large parallel momenta, where the topologically protected surface state is expected to resonate with the edge of the bulk conduction band [13] and to lose its spin-momentum locking character [30], was never accessed previously with trARPES.

In this work, we close this gap by performing high-resolution photoemission spectroscopy measurements of the occupied and unoccupied surface state manifold of a $Bi_2Se_3$(0001) surface up to 1.1 eV from the Dirac point and ±0.5 Å$^{-1}$ in the parallel momenta. The experiments are performed with an apparatus that couples a widely-tunable pump-probe beamline with an innovative ARPES detector: a photoemission electron microscope (PEEM) operating in momentum mode [31], also called momentum microscope (MM). With its ultimate angular acceptance, this instrument allows the acquisition of the entire photoemission horizon generated by a 6 eV femtosecond-laser probe, merging the high-resolution capabilities of low-photon-energy ARPES [32] with the wide momentum field-of-view typical of higher energy photon sources. We compare our findings with state-of-the-art theoretical predictions of the $Bi_2Se_3$ (0001) surface spectrum [33] and find a remarkable agreement between the calculated and the experimental quasiparticle spectrum when the GW approximation is employed.

## II. Experimental

The experimental set-up develops upon two commercially available sub-systems: a computer-controlled, turn-key, femtosecond laser source from Light Conversion and a KREIOS $^{TM}$ PEEM manufactured by Specs GmbH. A simplified diagram is reported in Fig.1.

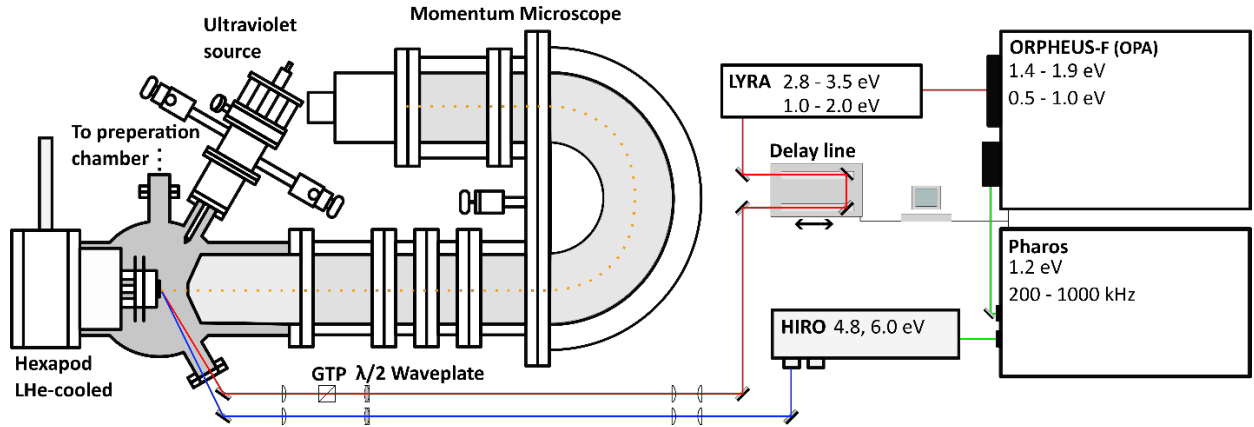

**Fig.1** Schematic diagram of the experimental setup.

The laser system is coupled to the photoemission end-station in a conventional pump-probe configuration. The primary light source is a 20 W, diode-pumped Ytterbium-based amplified laser emitting nearly Fourier-transform-limited femtosecond pulses with a central wavelength of 1028 nm and a time duration of 300 fs FWHM (Pharos $^{TM}$). This laser source allows on-demand selection of the pulse repetition rate up to 1 MHz. A substantial fraction of the output beam is fed to an optical parametric amplifier (OPA) (Orpheus-F $^{TM}$) to generate the optical pump employed in the time-resolved experiments. An ancillary second harmonic generation unit (Lyra

TM) extends the tuning range of the system to the near-UV wavelengths allowing an almost seamless selection of the pump photon energy between 0.5 eV and 3.5 eV with micro-joule-level pulse energies at 208 kHz [34].

A fifth harmonic generation stage (Hiro TM) converts a fraction of the fundamental beam into the 6 eV probe. Thanks to its high conversion efficiency (≈1% at 208 kHz), the fifth harmonic stage accepts a wide range of pulse energies at its input and can operate at high repetition rates. We exploit this feature in static photoemission experiments to increase the photoelectron count rate without exacerbating the space charge effects [35,36]. In fact, in the static mode, the constraint imposed by the OPA operation is lifted, and by setting a 1MHz repetition rate, an almost five-fold increase in the photoemission intensity is achieved without variations in the photoelectron-cloud density. The repetition rate can be switched through a computer-controlled interface without requiring changes in the optical system.

In the PEEM instrument, an immersion lens column generates an image of the lateral distribution of the photoelectrons ($x,y$) and an additional Fourier lens transforms it into an image of the photoelectron emission angles. The latter contains the momentum distribution of the photoelectrons $I(k_x,k_y)$ with high angular resolution [37]. The photoelectrons are filtered by an hemispherical-analyzer section of the instrument, which thus acquires momentum-resolved two-dimensional (2D) maps at a fixed, and selectable, kinetic energy ($E,k_x,k_y$).

Clean $Bi_2Se_3$(0001) surfaces are prepared in-situ by cleaving a commercially available crystal (2D Semiconductors) under high vacuum conditions (base pressure ~$10^{-8}$ mbar) and are transferred to the analysis chamber (base pressure < $1 \times 10^{-10}$ mbar) without additional treatments. The sample is cooled down with liquid helium throughout all the measurements, reaching an estimated temperature of 8 K.

Both the ARPES and trARPES spectra are collected with a pass energy of 50 eV and the narrowest slit (0.2 mm) at the entrance of the hemispherical analyzer. A field aperture is introduced in the intermediate image plane of the microscope column to restrict the area of the sample contributing to the signal (diameter: 70 µm) and select a uniform region of the surface. Due to the low photon energy employed for the pump pulse, which restricts the range of accessible parallel momenta, and to the limited extent of the surface state in the reciprocal space, the instrument is operated with high momentum-magnification factors, which corresponds to a k field-of-view of either ±0.5 Å$^{-1}$ (MAG.4) or ±0.2 Å$^{-1}$ (MAG.5). The momentum coordinates are calibrated by fitting the boundaries of the photoemission horizon with the free-electron dispersion relation.

The mapping of the occupied states band structure is performed by illuminating the sample with the p-polarized 6 eV probe only and at a repetition rate of 1 MHz. In the measurements of the unoccupied states band structure and of the photoexcited carrier dynamics instead, the repetition rate of the laser system is set to 208 kHz. The pump-photon energy is tuned to 0.97 eV and its fluence set to 310 µJ/cm$^2$. Both the pump and the probe beam are p-polarized. Being interested in the unoccupied electronic structure at the edges of the conduction band the photon energy is chosen to be resonant with a weak optical absorption edge found in Bi$_2$Se$_3$ [38] and assigned to a valence to conduction band transition [38,39]. This low photon-energy also suppresses spurious multiphoton photoemission from the pump beam due to the large work-function of the sample, which is found to be in the order of 5.2 eV. As a result, the nonlinear-photoemission background from the pump beam alone is below the dark-count rate of the photoelectron detector throughout the experiments.

## III. Results and discussion

### III.a Occupied and unoccupied electronic structure of Bi$_2$Se$_3$

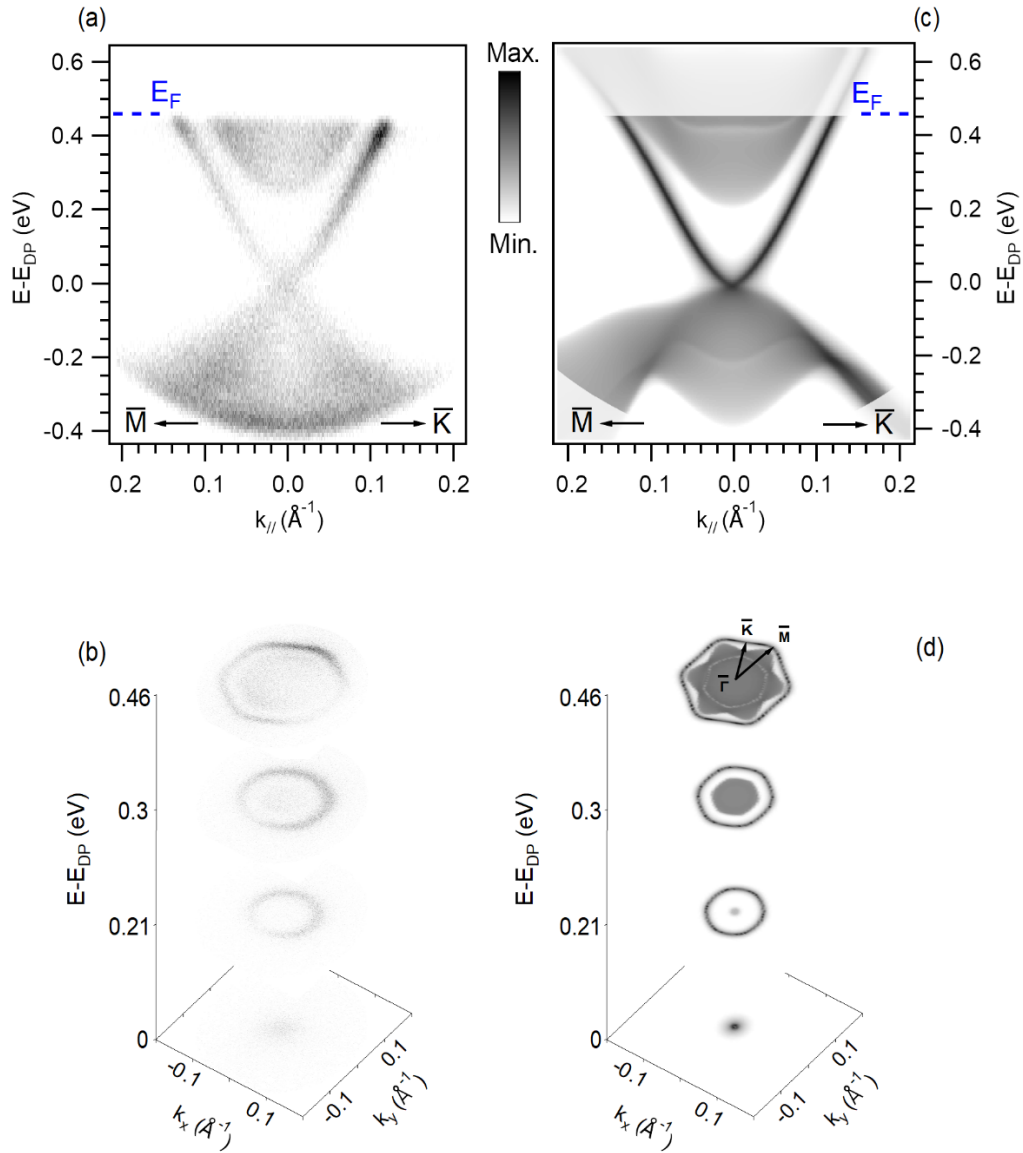

Figure 2 Photoemission spectra of Bi$_2$Se$_3$ collected at MAG.5 a) ARPES map along the $\overline{M}$-$\overline{\Gamma}$-$\overline{K}$ directions. b) Momentum distribution maps at four exemplificative energies above the Dirac point, E-E$_{DP}$. c) Theoretical surface spectrum calculated from the imaginary part of the surface Green function of a semi-infinite Bi$_2$Se$_3$ (0001) slab. The regions outside the photoemission horizon and above the equilibrium chemical potential of the sample are shaded for visualization purposes. d) Theoretical surface spectrum momentum distribution maps calculated at the energies of Fig.2b.

Before addressing the unoccupied electronic structure, we perform a conventional (static) ARPES characterization of the $Bi_2Se_3$(0001) surface. The results are reported in Fig.2.

The ARPES map in Fig.2a clearly shows the topologically-protected surface state[40] [13]of $Bi_2Se_3$, which manifests as linearly dispersing bands in the energy gap between valence and conduction bulk states. In this sample the Dirac point (DP), where ideally the density of states of the surface state should vanish, falls 0.46 ± 0.05 eV below the Fermi edge ($E_F$) and is taken as the origin of the energy scale. The presence of occupied bulk conduction-band states, revealed as the parabolic feature encompassed by the linear dispersing bands between 0.26 and 0.46 eV, highlights the intrinsic, degenerate, *n*-doping of the sample, which is believed to stem from defects in the crystal structure, such as selenium vacancies [41]. In bismuth selenide the doping level is also influenced by the adsorption of contaminants, which induces a strong band-bending and, eventually, leads to the emergence of a quantum-confined two-dimensional electron gas at the surface (2DEG) [42][43]. This phenomenon drives significant variations of work function and in the state-binding-energies as the adsorbates accumulate over the sample surface. In order to mitigate these drawbacks, we performed measurements on highly doped samples, where the effect of additional contamination is less severe [42] and refer the energy scale to the Dirac point to facilitate the comparison among the datasets and the theoretical predictions. The momentum distribution maps (Fig.2b) collected above the Dirac point highlight the well-known hexagonal warping of the surface state: the ring-like feature of the unperturbed surface state (E-$E_{DP}$= 0.21 eV) evolves into a hexagonal pattern as it approaches the Fermi level of the doped semiconductor. This phenomenon, which is due to the three-fold rotational symmetry of the surface [44] [42], is more evident where the surface state branches become degenerate with the bulk states ( E-$E_{DP}$= 0.3 eV and 0.46 eV). The corners of the hexagonal features fall along the $\bar{\Gamma}$-$\bar{M}$ directions and are exploited to determine the crystal orientation throughout the experiments. Our observation agrees with numerous precedent ARPES measurements, but, despite these results being established nowadays, the intriguing properties of topological matter are still driving the development of new approaches to rationalize the experimental evidences. One of such approaches, proposed by Aguilera *et al.* [33], makes use of many-body corrected tight-binding Hamiltonians generated for topological insulators of the $Bi_2Se_3$ family. The advantage of such an approach is the possibility of indirectly introducing the exchange and correlation effects, obtained with the GW method [45], to the electronic structure calculations of thick slabs and semi-infinite systems. Such an endeavor is beyond the current computational capabilities when DFT or GW methods are applied explicitly, while it can be tackled with a tight-binding approach employing the parameters obtained from DFT or GW calculations performed on a bulk system. Fig.2c and Fig.2d show the surface spectral function, *i.e.* the imaginary part of the surface Green function, of a $Bi_2Se_3$(0001) semi-infinite slab obtained through this approach. For a better comparison, it is displayed in the same energy and momentum regions probed by our experiment. The calculations are performed with the iterative Green-function method, as implemented in the software package WannierTools [46], starting from the real-valued Hamiltonian of $Bi_2Se_3$ published in ref.[33]. The excellent agreement between the experimental (Fig2.a) data and the theoretical prediction (Fig.2c) is evident in the dispersion of the surface state along the $\overline{M}$-$\overline{\Gamma}$-$\overline{K}$ directions, while a small deviation in the onset of the conduction band edge is observed. This mismatch is in the order of 50 meV and is comparable to the energy resolution of the instrument (see Supplementary Information, section S1). Therefore, it can be related to the uncertainty in the Dirac point energy, which we extracted from the data. The presence of a 2DEG on the surface [42] could also contribute to the observed deviation as the effects of the electron-doping are not included in the calculations. The overall agreement between the experimental and theoretical predictions is also further highlighted by Fig.2d, that reports calculated momentum distribution maps to be compared to the experimental ones in Fig.2b.

Having characterized our sample in the static photoemission regime, we now address its unoccupied states manifold. To this end, we illuminate the surface with a 0.97 eV pump beam at a fluence of 310 µJ/cm$^2$ (see experimental section). Upon the arrival of the pump pulse, an electronic population is promoted to both the unoccupied part of the surface states and the bulk conduction band. We acquire an energy-momentum space snapshot of this out-of-equilibrium population 0.77 ps after the zero-delay condition between the pump and the probe pulses to ensure the complete thermalization of the electron distribution [21], to avoid any effect induced by the temporal overlap [47], and to archive a statistical occupation of all the states within our measurement window (See Supplementary Information).

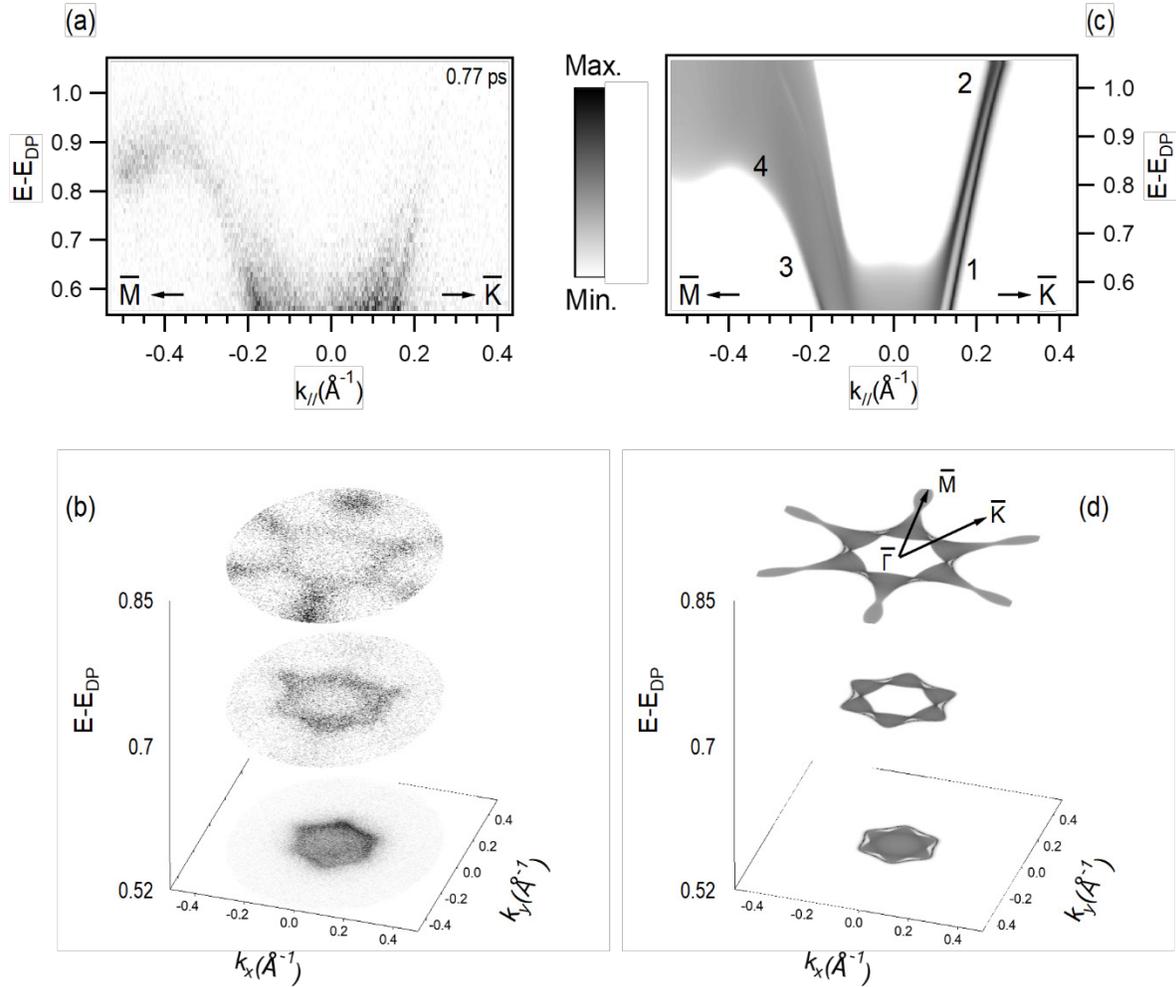

**Figure 3** Energy and momentum resolved spectra of the unoccupied states in Bi$_2$Se$_3$ collected at MAG.4 a) ARPES map along the $\overline{M}$-$\overline{\Gamma}$-$\overline{K}$ directions 0.77 ps after the optical excitation. b) Momentum distribution maps of the unoccupied states at three selected energies above the Dirac point, E-E$_{DP}$. c) Theoretical surface spectrum from the imaginary part of the Green function of a semi-infinite Bi$_2$Se$_3$(0001) slab. d) Theoretical surface spectrum momentum distribution maps calculated at the energies of Fig.3b.

Our findings, together with the theoretical predictions based on the approach introduced in the previous paragraphs, are summarized in Fig.3. Fig.3a shows the ARPES map (along the $\overline{M}$-$\overline{\Gamma}$-$\overline{K}$ directions) of the transiently-occupied states located from 0.5 eV to 1.1 eV above the energy of the Dirac point. The linear-dispersing signature of the surface state continues, almost unaltered, along the $\overline{\Gamma}$-$\overline{K}$ direction and stays visible up to an energy E-E$_{DP}$ of 0.8 eV. Above this energy, the photoemission intensity drops, and only a diffused background extends up to the edge of the measurement window. Along the $\overline{\Gamma}$-$\overline{M}$ direction, the signature of the surface state is more evident but deviates from the linear dispersion and evolves in a local maximum at 0.37 Å$^{-1}$. At the limit of the momentum window, 0.49 Å$^{-1}$, a local minimum can also be discerned. The structure of the unoccupied state manifold is further highlighted in Fig.3b where the momentum distribution maps, collected at three selected energies from the Dirac point, are reported. The three-fold rotational symmetry of the surface is evident in all the maps, and the distortion due to the hexagonal warping develops into a six-pointed-star shaped feature as the intermediate state energy increases. Fig.3c and Fig3d display the theoretical ARPES map and the theoretical momentum distributions maps, respectively. A simple visual comparison reveals how the theoretical model reasonably grasps the experimental observations. In particular, we observe a remarkable agreement between the experimental data and the theoretical predictions along the $\bar{\Gamma}$-$\bar{M}$ direction, where the local maxima and minima sequences well match in both the energy and the momentum coordinates. The agreement is less obvious along the $\bar{\Gamma}$-$\bar{K}$ direction, as the intensity of the photoemission signal drops quickly above E-$E_{DP}$ = 0.8 eV. However, below this energy, two linearly dispersing features are discernible.

In ref.[33], a tight- binding simulation performed on a 100 quintuple-layers slab of $Bi_2Se_3$(0001), employing the same many-body corrected Hamiltonian, is reported. It highlights the degree of localization of the states observed in our energy-momentum window. Following these results, we assign the feature labeled as 1 in Fig.3c as the true surface state, which is predicted to retain a strong localization on the first quintuple layer up to energies above 1 eV from the Dirac point. Feature 2 (Fig.3c) has a bulk-conduction-band nature with some minor contributions from surface resonances. Along the $\bar{\Gamma}$-$\bar{M}$ direction, the situation is dramatically different as the surface state quickly evolves into a surface resonance (feature 3 Fig3.c) and merges into an edge of the bulk conduction band (feature 4 Fig3.c) already at 0.7 eV from the Dirac point, losing its surface localization. Some of the differences observed between the data and the theoretical predictions, such as the drop in the intensity of the surface state along the $\bar{\Gamma}$-$\bar{K}$ direction, may be rationalized in terms of matrix-element or final-state effects. The latter are, in general, important in low-photon-energy photoemission as the final state cannot be approximated by the nearly-free-electron model.

A good agreement between GW predictions and ARPES data is not necessarily surprising. However, here, we observe a correspondence between the theoretical surface spectrum and the ARPES distribution of an out-of-equilibrium system, whose spectral function was perturbed by a femtosecond optical excitation prior to the electron-removal step. This correspondence implies a small electron-hole interaction, *i.e.* negligible excitonic effects, that is likely screened by the background electron population in the *n*-doped sample. Moreover, the results of Fig.3c,d are not obtained through an explicit GW calculation of the surface but indirectly introduced as correction in a parametrized tight-binding Hamiltonian on a Wannier functions basis [33]. Therefore, our results testify to the validity of this computationally less expensive approach in approximating the effects of many-body interactions on the $Bi_2Se_3$ surface.

### III.b Picosecond dynamics of hot electrons in $Bi_2Se_3$

The ultrafast electron dynamics of $Bi_2Se_3$ has already been the subject of numerous tr-ARPES investigations [11,20–25,48]. A coherent picture of the basic underlying processes emerges from these studies: depending on the doping of the sample, and on the photon energy, upon the arrival of the pump pulse interband and intraband transitions populate both the surface state and the conduction band states above their equilibrium chemical potential. Irrespectively on the transition pathway the population thermalizes, *via* electron-electron scattering, to a Fermi-Dirac distribution with a high temperature and a new effective chemical potential within 40 fs [21,48,49] from the initial excitation. The excess energy is subsequently dissipated to the lattice, on a picosecond time scale, through electron-phonon scattering and eventually diffuses away from the photo-excited region restoring the original equilibrium condition. In the present study, the 0.97eV optical excitation is expected to couple valence to conduction band states [38,39] and we address the cooling dynamics of the hot electron distribution in the whole energy-momentum region explored in section III.a. The experimental data and the details of the analysis are reported in the Supplementary Information (Section S1). In particular, by following a procedure similar to the one introduced in ref.[22] we quantify the temperature and the transient chemical potential of the electron population in the topologically-protected surface state and in the conduction band resonances. Upon the optical excitation the electronic temperature increases to about 900 K and is accompanied by a transient reduction of the chemical potential in the order of 30 meV. The electronic temperature relaxes with an exponential time constant of $\tau=3.57\pm0.09$ ps, revealing an electron-phonon scattering rate of $0.285\pm0.007$ THz, in good agreement with previous investigations of highly *n*-doped samples [22]. The negative variation of the chemical potential suggests that the interband transition at 0.97 eV is unable to significantly alter the population in the Dirac cone, despite the surface-resonance nature of the states along the $\bar{\Gamma}$-$\bar{M}$ direction. Moreover, despite their large excess energy with respect to the equilibrium chemical potential the surface resonances present long population dynamics, in the order of 500 fs.

### IV. Conclusions

We have mapped the band dispersion of the unoccupied states in a prototypical topological insulator up to 1.1 eV from the Dirac point and 0.5 Å$^{-1}$ in parallel momenta, close to the boundary of the photoemission horizon from the 6 eV probe. Our measurements highlight the momentum-dependent evolution of the Dirac cone into surface resonances at the edge of the conduction band, as predicted by a state-of-the-art theoretical model of the surface's electronic structure. The necessity to probe wide energy-momentum volumes in time-resolved experiments is one of the driving forces behind the rapid diffusion of trARPES set-ups based on femtosecond extreme ultraviolet light sources, which can illuminate energy-momentum regions inaccessible to low-photon-energy ARPES. Despite the advantages of these last-generation systems, we argue that combining a momentum microscopy apparatus with a conventional laser-based source is a viable approach to probe the out-of-equilibrium electronic structure in a wide momentum field of view, covering a large portion of the surface Brillouin zone of many crystalline materials.

**Acknowledgments**

We acknowledge funding from the DFG (Major Research Instrumentation Individual Proposal INST 212/409-1), the Ministerium für Kultur und Wissenschaft des Landes Nordrhein-Westfalen (NRW), and from the European Research Council (ERC) under the European Union's Horizon 2020 research and innovation programme (Project hyControl, grant agreement No 725767).

# Supplementary Information: Dirac bands in the topological insulator Bi$_2$Se$_3$ mapped by time-resolved momentum microscopy

Stefano Ponzoni[1,*,§], Felix Paßlack[1, §], Matija Stupar[1], David Maximilian Janas[1], Giovanni Zamborlini[1,*] and Mirko Cinchetti[1].

[1] Department of Physics, TU Dortmund University, 44221 Dortmund, Germany

§ These authors contributed equally to this work.
*Corresponding authors:
stefano.ponzoni@tu-dortmund.de; giovanni.zamborlini@tu-dortmund.de

**S1. Picosecond dynamics of hot electrons in Bi$_2$Se$_3$(0001)**

The ultrafast electron dynamics of Bi$_2$Se$_3$ has already been extensively investigated with both two-photon photoemission and tr-ARPES measurements [1–8]. Most of the previous studies focused on its dependence with the doping level, on the excitation density, or on the equilibrium temperature of the lattice and were performed with a 1.5 eV excitation. At this pump-photon-energy the optical transition in *p*-doped Bi$_2$Se$_3$ mainly involves valence and conduction bulk states [6], while for *n*-doped samples, such as the one used in this work, an additional absorption pathway opens and the bottom edge of the conduction band is coupled with a second topologically-protected surface state located in a higher energy gap of the surface-projected band structure [4,8]. Irrespectively on the transition pathway the population of the Dirac cone thermalizes, *via* electron-electron scattering, to a Fermi-Dirac distribution with a high temperature and a new effective chemical potential within 40 fs [2,8,9] from the initial excitation. In our study, we reduce the pump photon energy from 1.5eV to 0.97eV, so that the optical excitation is expected to couple only valence to conduction band states [10,11] (Fig. 1S) and we address the hot electron cooling dynamics in the whole energy-momentum region explored in section III.a of the main text.

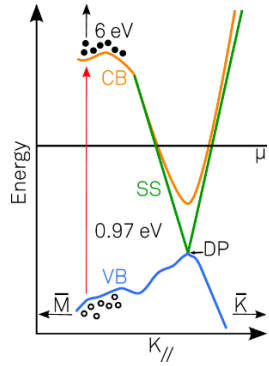

Fig.1S Proposed scheme of the optical transition upon 0.97 eV laser excitation. The pump photons couple the states ad the edge of the valence band to conduction band states along the $\bar{\Gamma}$-$\bar{G}$ direction [11].

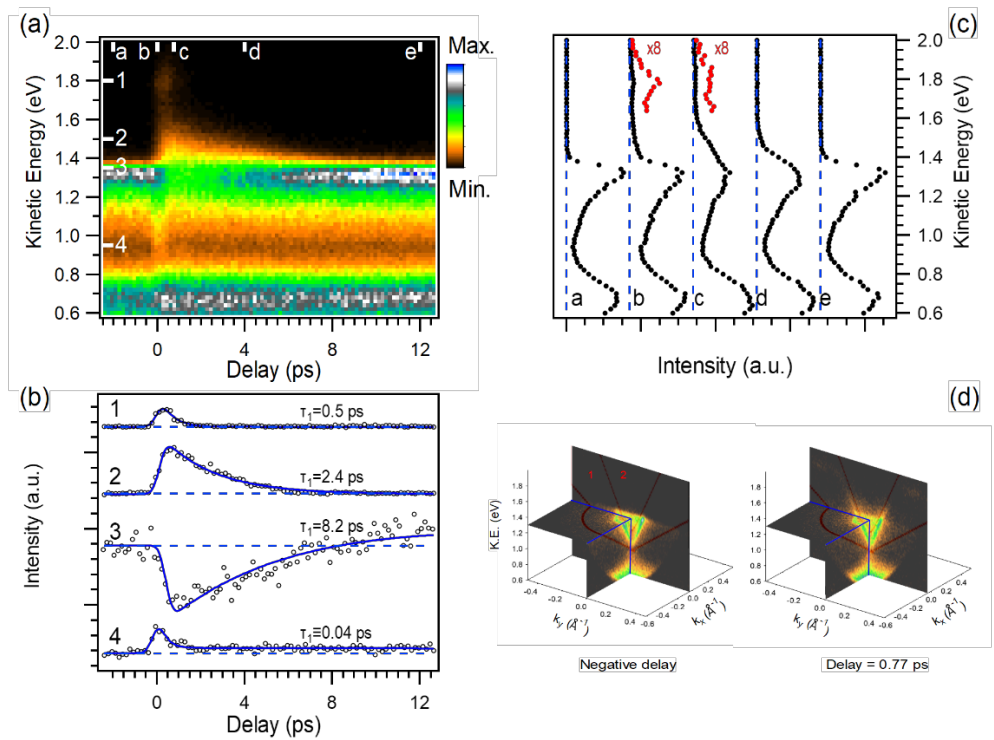

Figure 2S. Quasiparticle dynamics in $Bi_2Se_3$. a) Momentum-integrated map. The markers from 1 to 4 and from a to e indicate the energy and the delay coordinates of the data slices in fig.2Sb and fig.2Sc. b) Time-resolved dynamics at 4 exemplificative kinetic energies (black circles). The best-fit results (blue line) and the corresponding time constants, $\tau_1$, are reported. c) Energy distribution curves at five exemplificative delays (black dots). Scaling the signal by a factor 8 highlights the variations at high intermediate-state-energy (Red dots). d) Exemplificative energy end momentum resolved spectra before and after the time time-coincidence. The red lines defines the boundary between the surface states dominated (1) and the bulk conduction band (2) regions used in the analysis. In this picture, several data slices for each intersection plane are averaged together for visualization purposes.

Fig.2S summarizes the typical results of a time resolved experiment. Here the Dirac point falls at a binding energy of 0.40 eV and the edge of the bulk conduction is about 190 meV below the equilibrium chemical potential (at kinetic energies of 0.98 eV and 1.17 eV, respectively). Since the equilibrium chemical potential is not a good energy reference for a hot electron distribution whose temperature evolves on a picosecond time scale [2], the data are displayed in function of the electron kinetic energy as given by the MM. In this scale the equilibrium chemical potential of the unexcited sample falls at 1.366±0.005 eV. This value, together with the energy resolution of the analyzer (50 meV), are determined from static ARPES experiments on the $Bi_2Se_3$ surfaces and on an Au(111) single crystal.

In fig.2Sa, the momentum-integrated energy- and time-resolved photoelectron distribution is displayed. The energy step is 10 meV and the delay step 166 fs. The dynamics of the photoemission signal is highlighted in Fig.2Sb, where time-resolved profiles are reported for a few selected kinetic energies (black circles). The corresponding profiles are also indicated with labels, from 1 to 4, in Fig.2Sa. In Fig. 2Sc the momentum-integrated EDC are displayed for a few selected delays. In analogy with the former plot the markers from *a* to *e* indicate the corresponding profile in the energy- and time- resolved map.

An initial, fast, transient appears in the energy range between 0.9-1.2 eV and is later followed by a depopulation (population) dynamics of the states below (above) the equilibrium chemical potential evolving on a multi-picosecond time scale. A step-like enhancement of the photoemission rate is found in the region between 0.6 eV and 1.1 eV where it emerges immediately after the laser excitation and persists in the whole delay window. Moreover, a feature with enhanced spectral weight can be discerned at about 1.8 eV near the time coincidence. This energy corresponds to the sequence of local critical points long the $\bar{\Gamma}$-$\bar{M}$ direction at the edge of the bulk valence band (Fig.3a, Fig.3c of the main text).

We assign the fast transient between 0.9 eV and 1.2 eV to an image potential state (IPS) excited by the 6 eV pulse and photoemitted by the 0.97 eV photons. The presence of an IPS on the $Bi_2Se_3$ (0001) surface has already been reported by several authors [1][8]. In particular, Sobota *et al.*[1], determined a binding energy of 0.77 eV with respect to the vacuum level by tr-ARPES measurements with a 6 eV and 1.5 eV photons. These parameters are similar to our results (B.E.$_{VL}$=0.59±0.05 eV and effective mass = 1.3±0.1 $m_e$ (see section S2).

Since the lifetime of the IPS is in the order of 40 fs [12], well below the duration of the laser pulses, we exploit its dynamics to estimate their cross-correlation. The signal at K.E= 0.95 eV (Fig.2Sb, profile 4 ), is fitted with the convolution of a gaussian term with the sum of an exponential and a step-like function. By setting the time constant to 40 fs we obtain a cross-correlation of 580±50 fs FWHM. The same phenomenological fitting function well approximates all the transients reported in fig.2Sb with the same cross-correlation value (blue lines). The best-fit time constants,

$\tau_1$, of the first exponential term are also reported in Fig.2S b, for clarity. While these multi-picosecond, quasi-exponential, energy-dependent relaxations [2] are the fingerprints of the well-known $Bi_2Se_3$ hot-electron cooling dynamics, the step-like transient was never reported in previous tr-ARPES studies. This dynamics, which is evident at all the kinetic energies below 1.2 eV, relaxes on a time scale much longer than our delay window. Addressing is physical origin requires further investigations and goes beyond the scope of this work.

Despite their bulk nature, and their large excess energy with respect to the equilibrium chemical potential, the states located at the edge of the conduction band display a population dynamics in the order of 500 fs (Fig.2Sb, profile 1). These states, being already present at zero delay (Fig.2Sb and Fig.2Sc), are among the firsts to be populated by the pump, it is therefore tempting to identify them as the final states of the optical transition, as originally suggested in Ref.[11]. Nevertheless, an unambiguous identification would require a time resolution below the electron-electron scattering time.

Following a procedure similar to the one introduced in ref.[3] we quantify the temperature and of the transient chemical potential of the electron population in the topologically-protected surface state and the conduction band resonances. We integrate the time resolved transient over two momentum volumes which separate the region dominated by the surface state contribution from the bottom edge of the bulk conduction band. The intersections of the regions' boundaries with the cut-planes are also displayed in Fig.2Sd, and the encompassed areas labeled 1 and 2, respectively. The volumes are defined from the equilibrium $I(E,k_x,k_y)$ distribution and applied unaltered for all the delays. Two exemplificative EDC curves, steaming from region 1 only, are reported in Fig.2Sa for the equilibrium condition (before delay zero, solid black markers) and at 2.5 ps after the optical excitation (solid red markers).

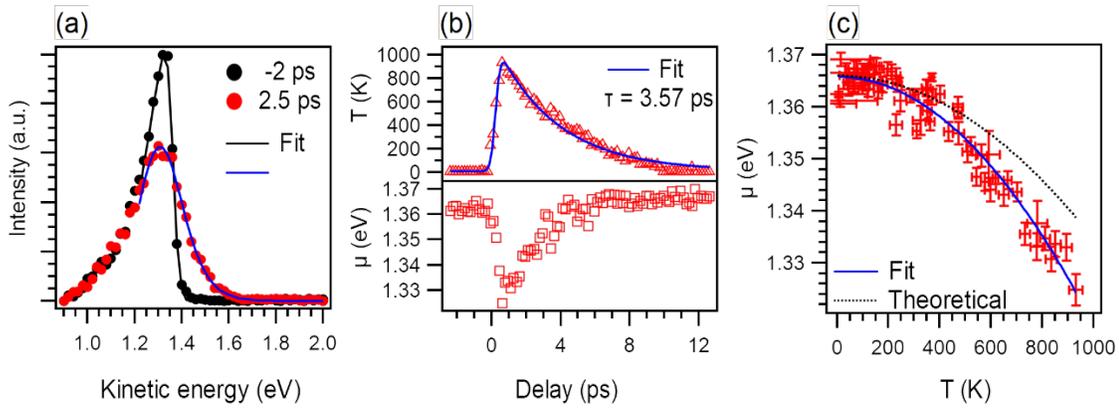

Figure 3S. Hot-electron cooling dynamics in the surface-state-dominated energy-momentum region. a) Energy distribution curve 2 ps before (black dots) and 2.5 ps after (red dots) the temporal coincidence. The best-fit results using Eq.1 are displayed (solid lines). b) Evolution of the electronic temperature (red triangles) and of the chemical potential (red square) as obtained from the fitting with Eq.1. The temperature-decay time, using a single exponential model, is τ =3.57 ps. c) Chemical potential versus electronic temperature. The blue line is the best-fit result using a quadratic model. The theoretical prediction (Black dotted line) is reported for comparison.

All the EDCs are fitted with the function:

$$I(E_k) = \int_{-\infty}^{+\infty} \frac{1}{e^{\left(\frac{E_k-\mu}{k_b T}\right)}+1} A(E_k - E_{DP}) G(E_k - \epsilon) d\epsilon \quad (1)$$

A convolution of a gaussian term $G(E_k - \epsilon)$, representing the energy resolution of the analyzer, with the linear density of states (DOS), $A(E_k - E_{DP})$, of the surface states weighted by the Fermi-Dirac distribution. In this equation $E_k$ is the kinetic energy of the photoelectrons, $E_{DP}$ is the position of the Dirac point and μ is the chemical potential, all expressed in the kinetic energy scale of the analyzer. The constant A encompass all the multiplicative terms which modulate the photoemission intensity, such as the photoemission matrix elements, the probe intensity and the transmission of the analyzer. It can be regarded as the effective "slope" of the linear DOS in the surface state. Its value is determined, together with the energy of the Dirac point ($E_{DP}$), by fitting of an average of the EDCs taken at delays prior to the excitation. The best-fit result is displayed in Fig.4a (Black line). The so-determined values of A and $E_{DP}$ are kept fixed for all the delays reducing the free fitting parameters to only µ and T. This fitting function is one of the simplest approximation of the transient signal and relies on many assumptions concerning the photoemission process, such as the invariance of the photoemission matrix elements for all the states probed in the experiment, a negligible time dependent energy-broadening of the transient states [13] [14] and no time-dependent variations of the overall photoemission intensity [15]. The latter two effects were taken into account in ref.[3], but we chose not to introduce them to limit the number of free parameters in our analysis. Despite the simplicity of the model, the out of equilibrium EDCs are generally well-fitted and an exemplificative result is reported in Fig.3Sa (Blue line). The temperature and the chemical potential obtained from the analysis are displayed in Fig.3Sb. Upon the optical excitation the electronic temperature increases to about 900 K and is accompanied by a transient reduction of the chemical potential in the order of 30 meV. The hot electron distribution relaxes with an exponential time constant of τ=3.57±0.09 ps, revealing an electron-phonon scattering rate of 0.280±0.007 THz, in good agreement with previous investigations of n-doped samples [3]. A strong temperature dependence of chemical potential is a characteristic of two-dimensional electron gases with linear energy-dispersion relations. Assuming the conservation of the number of particles, $n$, and by taking the Sommerfeld expansion of the integral $n = \int_0^\infty fd(E,T)DOS(E)dE$ (where $fd(E,T)$ is the Fermi-Dirac distribution) up to the second order in the temperature, the expression of the chemical potential results in:

$$\mu(T) = E_f - \frac{\pi^2 k_b^2}{6E_f}T^2 \quad (2)$$

In Eq.2 $E_f$ Is the fermi energy relative to the Dirac point, which is considered as the origin of the band. The variation is independent on the slope of the band and is significatively larger than for a three dimensional system with quadratic band dispersion[16]. In Fig.3Sc we compare this result to the experimental data. The red markers are the value of the chemical potential in function of the temperature as extracted by the fitting procedure (Eq.1). The error bars correspond to the statistical uncertainty given by the analysis software, and were omitted in graph of Fig.3Sb for clarity. The best-fit with a quadratic relation (red line) gives a temperature coefficient of (-4.7±0.1)X$10^{-8}$, which deviates from the theoretical value (-3,15X$10^{-8}$) by 49 %. The theoretical variation in the chemical potential is plotted in Fig3S.c as a black dotted line for comparison. Despite the mismatch, the trend we observe is in the same order of the prediction and similar to the results reported in a previous investigation with 1.5 eV pump-photons [3]. Eq.2 implies the conservation of the number of particles in the linear dispersing states, therefore our results suggest that the interband optical excitation among bulk states with 0.97 eV photons is unable to significatively alter the population and the charge balance of the Dirac cone. In particular, an increase in the number of particles due to bulk-to-surface scattering processes would lead to a positive variation of the chemical potential, opposite to our observation. Other investigators indeed reported such positive variations on a multi-picosecond time-scale [2] or in a sub-picosecond interval close to the optical excitation [3], but always after an optical transition induced by 1.5 eV photons.

## S2. Image potential state of $Bi_2Se_3(0001)$

In order to identify the origin of the fast transient observed in the momentum integrated time-resolved spectrum (Fig.2Sa, 2Sb) we collected ARPES data-cubes at its maximum signal-intensity. The ARPES maps along the $\overline{M}$-$\overline{\Gamma}$-$\overline{M}$ and the $\overline{K}$-$\overline{\Gamma}$-$\overline{K}$ directions are displayed in fig 4S.a and 4S.b, respectively.

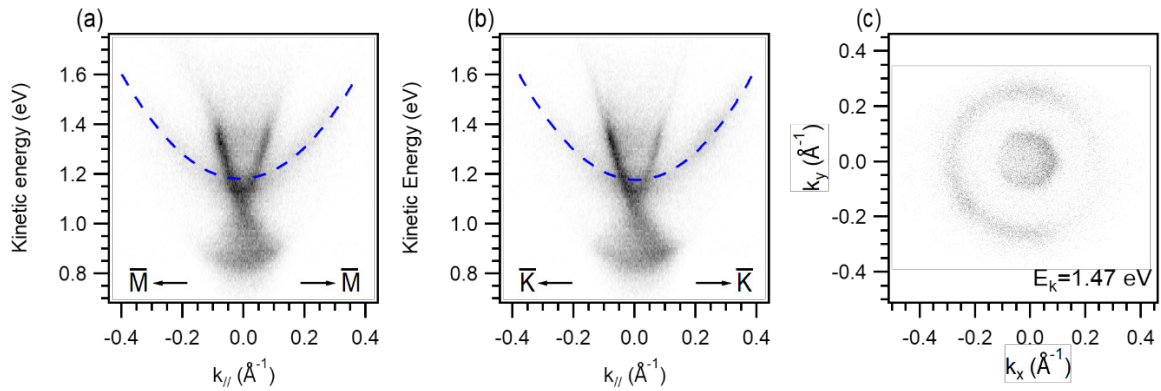

Figure 4S: ARPES measurements of the IPS state of $Bi_2Se_3$. a) ARPES map along $\overline{M}$-$\overline{\Gamma}$-$\overline{M}$ direction. b) ARPES map along the $\overline{K}$-$\overline{\Gamma}$-$\overline{K}$ direction. c) Momentum distribution map collected at a kinetic energy of 1.47 eV.

At this delay, an additional band, with parabolic energy-momentum dispersion, appears in the ARPES maps. Its dispersion is independent from the high symmetry direction and is well-fitted using a nearly-free electron model with an effective of mass of 1.3±0.1 $m_e$ (Fig.4Sa,b blue dashed line). Its rotational symmetry is highlighted in the momentum distribution map of Fig.4Sc, which is averaged in a narrow energy window around 1.47 eV (in the kinetic energy scale of the analyzer). By comparison, the hexagonal warping of the unoccupied surface state is evident. These observation are consistent with the properties of an image potential state (IPS): an electron bound in front of a polarizable surface by the effect of its own image charge. [17]

As discussed in the main text, the excitation of IPS by 6 eV radiation has already been reported on $Bi_2Se_3$ surfaces.[1 8 12] Given the photon-energy of the pump pulse, which here is responsible to the ionization of the IPS, and the work function of the sample (5.43±0.05 eV for the data of Fig.4S

) we obtain a binding energy of 0.59±0.05 eV with respect to the vacuum level. This value deviates by 180 meV from the results reported by Sobota *et al.*[1] This mismatch could arise from a number of factors including, for example, the presence of contaminants adsorbed on the surface, which are known to alter the properties of the IPS.[18] Since the parabolic dispersion of the state can be followed to more than 0.3 Å$^{-1}$, alternative explanations, such as laser assisted photoemission phenomena[19] can be ruled out: no bulk or surface state with nearly-free character are expected in $Bi_2Se_3$ in the energy and momentum window probed by the experiment.[20][21] As a final note, we underline that the data reported in Fig.4S were collected on a sample with a lower level of doping than the ones used in the main text. The consequent variations in work function and in the kinetic energies at which the states fall are evident. The situation for the IPS is no different: being pinned to the vacuum level, its minimum shifts together with the secondary edge, and appears at higher kinetic energies with respect to the transients reported in Fig.2S.